\documentclass{iopconfser}
\usepackage[]{graphicx}
\usepackage[]{caption}
\usepackage[]{subcaption}
\usepackage{cite}
\usepackage{pifont}
\usepackage[]{amsmath}
\begin{document}

	\title{Linearly Interacting Barrow Holographic Dark Energy in Modified Cosmology}
	
	\author{Suphakorn Chunlen$^{* 1}$ and Kraiwut Dakam$^{** 1}$}
	
	\affil{$^1$Department of Physics, School of Science, University of Phayao, Phayao, Thailand}
	
	\email{suphakorn.official@gmail.com$^*$}
	\email{64202947@up.ac.th$^{**}$ (corresponding author)}
	
	\begin{abstract}
		We study the Barrow holographic dark energy (BHDE) model using the apparent horizon as the infrared (IR) cut-off and ensuring thermodynamic consistency. However, the results show that the BHDE model without any interaction does not yield the accelerated expansion of the universe. Including a linear interaction solves the acceleration and dark energy equation of state parameter problems.
	\end{abstract}

\section{Introduction}
\hspace{0.5cm} According to observable cosmological parameters \cite{aghanim2020planck}, there is evidence that the present universe is accelerating in its expansion. Physicists have been constructing many theoretical models to explain this phenomenon. One of the many various models is the dark energy model. The most acceptable model is the $\Lambda CDM$ model which has given consistent results with observable cosmological data, but it cannot answer the cosmological constant problem \cite{peebles2003cosmological,copeland2006dynamics}. Recently, the holographic dark energy (HDE) based on the holographic principle have been constructed to investigate the quantum fluctuations of spacetime, which depend on the universe boundary \cite{gao2013explaining,susskind1995world} at a distance of the cosmological length scale or the IR cut-off $L$. Many types of the IR cut-off have been considered in the literature \cite{li2004model,Cruz:2018lcx,Sharma:2020glf,Nojiri:2005pu,Granda:2008dk,Chunlen:2020nru}, such as the particle horizon, the Hubble horizon and the future horizon.
According to the holographic principle, information in three-dimensional spacetime is imprinted on a two-dimensional surface \cite{bousso2002holographic,stephens1994black}. The standard HDE model has arisen from considering the Schwarzchild black hole entropy or the Bekenstein-Hawking entropy. This model with the Hubble horizon as the IR cutoff results in pressureless dark energy \cite{li2004model}. Furthermore, the model's equation of state $p_D = \omega_D \rho_D$, with $\omega_D = 0$, cannot lead to the accelerating expansion of the universe at the present time \cite{li2004model}. A way to solve the problem is to consider a variation of the model. In this work, we choose a type of HDE models where the energy density derived from the Barrow entropy, and this HDE model is called Barrow HDE or BHDE \cite{saridakis2020barrow}. However, this model with the Hubble horizon as the IR cut-off is unstable despite the fact that it yields the universe in the quintessence regime \cite{srivastava2021barrow}. Nonetheless, the result in \cite{aghanim2020planck} suggested that present phantom phase of the universe is favoured ($\omega_{D0} < -1$). Hence, in our study of the BHDE, we will use another choice of the IR cut-off that is in particular the apparent horizon to explore the model further. This paper is organised as follows. First, we investigate the Barrow HDE model with the apparent horizon with the thermodynamic consistency, which associates with the first and the second laws of thermodynamics \cite{sheykhi2021barrow,akbar2007thermodynamic,saridakis2020modified}. Next, we consider adding the general linear interaction in the dark sector where the interaction term we use has been considered in \cite{luciano2023generalized,wang2007interacting,wang2024further,amendola2000coupled,zimdahl2001}. Then we show the result and discussion of the cosmological parameters, explaining the universe expansion, the phase of the dark energy, and also the instability of the model. Finally, we conclude our findings.

\section{A Brief Overview of Barrow Holographic Dark Energy}
\hspace{0.5cm} The application of the holographic principle in the cosmological framework involves defining the boundary of the cosmological horizon entropy. Recently, Barrow pointed out that quantum gravitational effects may produce complex fractal features in the structure of black holes \cite{barrow2020area}, leading to a deformation of the expression for black hole entropy.

\begin{eqnarray}
	S = \left(\frac{A}{A_0}\right)^{\frac{2+\Delta}{2}}, \label{Barrow entropy}
\end{eqnarray}
where $A$ is the standard horizon area and $A_0$ is the Planck area. Barrow proposed that the Barrow exponent $\Delta$ quantifies the quantum-gravitational deformation, with $\Delta = 0$ corresponding to the simplest horizon structure as Bekenstein-Hawking entropy and $\Delta = 1$ corresponding to the most intricate and fractal structure \cite{saridakis2020barrow}. The deformation effects lead to different cosmological behaviour from the standard HDE. The standard holographic dark energy is given by the inequality $\rho_{D}L^4 \le S$ which relates to $S \propto A \propto L^2$, $L$ denotes as horizon length \cite{wang2017holographic}. Using the Barrow entropy (\ref{Barrow entropy}) leads to

\begin{eqnarray}
	\rho_D = CL^{\Delta-2}\label{BHDE}.
\end{eqnarray}
\hspace{0.5cm} This is the Barrow holographic dark energy with $C$ as a constant determined numerically, we consider a flat FLRW spacetime metric (using natural units) with constant curvature $k$:

\begin{eqnarray}
	ds^2 = -dt^2 + a^2\left(\frac{1}{1 - kr^2}dr^2 + r^2d\theta^2 + r^2\sin^2\theta d\phi^2\right),
\end{eqnarray}
and consider the largest horizon length as the apparent horizon:

\begin{eqnarray}
	\widetilde{r}_A = \frac{1}{\sqrt{H^2 + \frac{k}{a^2}}}\label{apparent horizon}.
\end{eqnarray}

\hspace{0.5cm} We consider the universe as a homogeneous and isotropic matter perfect fluid, as well as the HDE. From a thermodynamic point of view, the apparent horizon is also a suitable horizon because the consistency with the first and second laws of thermodynamics \cite{sheykhi2021barrow}. Moreover, we can relate the temperature to the apparent horizon defined as \cite{akbar2007thermodynamic, saridakis2020modified}

\begin{eqnarray}
	T_h = \frac{1}{2\pi\widetilde{r}_A}\left(1-\frac{\dot{\widetilde{r}}_A}{2H\widetilde{r}_A}\right).
\end{eqnarray}
\hspace{0.5cm} For the case with an infinitesimal internal of time, $\dot{\widetilde{r}}_A \ll 2H\widetilde{r}_A$, leads to defining $T_h = \frac{1}{2\pi\widetilde{r}_A}$. In the form of the perfect fluid of the matter and energy contents in the universe, the energy-momentum tensor is given by

\begin{eqnarray}
	T_{\mu\nu} = (\rho + p)u_\mu u_\nu + pg_{\mu\nu}, \label{energy-momentum tensor}
\end{eqnarray}
where $\rho$ and $p$ are the energy density and pressure, respectively. In this work, we propose that the total energy contents of the universe is considered by the conservation equation $\nabla^\mu T_{\mu\nu} = 0$, yielding

\begin{eqnarray}
	\dot{\rho} + 3H(p + \rho) = 0. \label{conservation equation}
\end{eqnarray}

\hspace{0.5cm} According to observable parameters, the universe is expanding with acceleration, meaning that we have a changing volume. For the FLRW metric spacetime as background with (\ref{energy-momentum tensor}), the work density is obtained as \cite{sheykhi2021barrow}

\begin{eqnarray}
	W = \frac{1}{2}(\rho - p).
\end{eqnarray}
\hspace{0.5cm} Then we use the first law of thermodynamics, which satisfies the apparent horizon, as

\begin{eqnarray}
	dE = T_h dS_h + W dV\label{the first law of thermodynamics},
\end{eqnarray}
where $E$ is the total energy of the universe enclosed within the apparent horizon, and $T_h$ and $S_h$ are the temperature and entropy, respectively, associated with that horizon. The last term of the first law of thermodynamics is the work density according to the accelerating expansion of the present universe.

\hspace{0.5cm} Taking the differential with respect to cosmic time of the three-dimensional spherical space with radius $\widetilde{r}_A$, assume that the volume of the universe within the surface area $A = 4\pi\widetilde{r}_A^2$ is $V = \frac{4\pi}{3}\widetilde{r}_A^3$ \cite{sheykhi2021barrow}. We immediately obtain

\begin{eqnarray}
	dE = 4\pi\widetilde{r}_A^2\rho d\widetilde{r}_A + \frac{4\pi}{3}\widetilde{r}_A^3\dot{\rho} dt,
\end{eqnarray}
and from the conservation equation (\ref{conservation equation}), we obtain

\begin{eqnarray}
	dE = 4\pi\widetilde{r}_A^2\rho d\widetilde{r}_A + 4\pi H\widetilde{r}_A^3(\rho + p) dt.
\end{eqnarray}

\hspace{0.5cm} To consider the universe with effects from the quantum-gravitational deformation, we use the Barrow entropy (\ref{Barrow entropy}) and associate it with the apparent horizon. In this step, we change the black hole event horizon radius to the apparent horizon and then take the differential form of the entropy. We obtain

\begin{eqnarray}
	dS_h = d\left(\frac{A}{A_0}\right)^{\frac{2+\Delta}{2}} = (2 + \Delta)\left(\frac{4\pi}{A_0}\right)^{\frac{2 + \Delta}{2}}\widetilde{r}_A^{1 + \Delta}\dot{\widetilde{r}}_Adt.
\end{eqnarray}

\hspace{0.5cm} By using the above equations and the first law of thermodynamics (\ref{the first law of thermodynamics}) and the radius of the apparent horizon (\ref{apparent horizon}) to find the differential form of the Friedmann equation

\begin{eqnarray}
	\frac{2 + \Delta}{2\pi A_0}\left(\frac{4\pi}{A_0}\right)^{\frac{\Delta}{2}}\frac{d\widetilde{r}_A}{\widetilde{r}_A^{3-\Delta}} = H(\rho + p)dt,
\end{eqnarray}

using the continuity equation (\ref{conservation equation}) after some calculation we find the modified Friedmann equation as \cite{sheykhi2021barrow}

\begin{eqnarray}
	\left(H^2 + \frac{k}{a^2}\right)^{1 - \frac{\Delta}{2}} = 8\pi G_{\rm{eff}}\frac{\rho}{3}\label{the modified Friedmann equation},
\end{eqnarray}

where Hubble parameter is $H \equiv \frac{\dot{a}}{a}$ and $G_{\rm{eff}}$ is defined as the effective Newtonian gravitational constant \cite{sheykhi2021barrow}

\begin{eqnarray}
	G_{\rm{eff}} = \frac{A_0}{4}\left(\frac{2 - \Delta}{2 + \Delta}\right)\left(\frac{A_0}{4\pi}\right)^{\frac{\Delta}{2}}
\end{eqnarray}
\section{The Linear Interaction of a Dark Sector in Barrow Holographic Dark Energy Model}
\hspace{0.5cm} In the context of cosmology, it is plausible to consider that the dark energy and dark matter sectors interact with each other. This interaction can be described through a coupling term in the energy conservation equations \cite{yang2023dynamics},
\begin{eqnarray}
	\nabla^\mu(T^{CDM}_{\mu\nu} + T^{DE}_{\mu\nu}) = 0.
\end{eqnarray}

\hspace{0.5cm} The energy conservation equations for dark energy and dark matter in the presence of an interaction term $Q$ are given by

\begin{eqnarray}
	\dot{\rho}_m + 3H\rho_m = Q, \label{DM conservation equation}\\
	\dot{\rho}_D + 3H(1 + \omega_D)\rho_D = -Q, \label{DE conservation equation}
\end{eqnarray}
where $\omega_D$ denotes as the dark energy equation of state parameter. In this framework, due to the unknown nature of dark energy and dark matter, we consider the most general linear interaction function, or the linearly dark sectors coupling interaction \cite{luciano2023generalized,wang2007interacting,wang2024further}, given by

\begin{eqnarray}
	Q = 3\alpha H\rho_m + 3\beta H\rho_D.
\end{eqnarray}
\hspace{0.5cm} This form of interaction function ensures that the interaction term is proportional to the energy densities of the dark sectors due to the expanding universe. For instance, a positive $Q$ indicates the dark energy decays into dark matter, which could help alleviate the coincidence problem \cite{amendola2000coupled}. Conversely, a negative of $Q$ suggests the dark matter decays into dark energy, which lead to different dynamics in the expansion of the universe \cite{zimdahl2001}. The effect of the coupling parameters $\alpha$ and $\beta$ on the behaviour of cosmological parameters is significant. These parameters influence the rate at which these interactions occur and can affect the evolution of the density parameters $\Omega_D$ and $\Omega_m$ over time. By analyzing the impact of these coupling constants, we can gain insights into the possible interactions between dark energy and dark matter and their role in the accelerating expansion of the universe.

\hspace{0.5cm} The modified Friedmann equations (\ref{the modified Friedmann equation}) can be converted into the equation of total density parameter as done in \cite{mohammadi2023friedmann}

\begin{eqnarray}
	\Omega_D + \Omega_m + \Omega_k = 1\label{OmegaAll},
\end{eqnarray}

we follow those steps to define the dimensionless density parameter of matter, curvature and dark energy as

\begin{eqnarray}
	\Omega_D = 8\pi G_{\rm{eff}}\frac{\rho_D}{3H^2}\left(H^2 + \frac{k}{a^2}\right)^{\frac{\Delta}{2}},\; \Omega_m = 8\pi G_{\rm{eff}}\frac{\rho_m}{3H^2}\left(H^2 + \frac{k}{a^2}\right)^{\frac{\Delta}{2}},\; \Omega_k = -\frac{k}{a^2H^2}\label{the density parameters}.
\end{eqnarray}

\hspace{0.5cm} Then we consider the conservation equations with the linear interaction (\ref{DM conservation equation}) and (\ref{DE conservation equation}), and the equation of state parameter is $\omega_D = \frac{p_D}{\rho_D}$. Differentiating (\ref{the modified Friedmann equation}) with respect to the scale factor and solving the equation by using (\ref{DM conservation equation}), (\ref{DE conservation equation}) and (\ref{the density parameters}), we can immediately obtain

\begin{eqnarray}
	\frac{H^\prime}{H} = \frac{-3(1 + \omega_D)\Omega_D + \Omega_m}{a(2 - \Delta)} - \frac{\Omega_k}{a}\label{diffFriedmann}.
\end{eqnarray}

\hspace{0.5cm} Here, a prime denotes as the derivative with respect to scale factor $a$. Substituting (\ref{BHDE}) with the length of horizon that is the apparent horizon (\ref{apparent horizon}) in (\ref{the modified Friedmann equation}) and use the initial condition at present where $a = a_0 = 1$, we are now able to evaluate the numerical constant

\begin{eqnarray}
	C = \frac{3\left[k + H_0^2(1 - \Omega_{m0})\right]}{8\pi G_{\rm{eff}}(H_0^2 + k)}\label{C}.
\end{eqnarray}

\hspace{0.5cm} By using the BHDE (\ref{BHDE}), dark energy conservation equation (\ref{DE conservation equation}) and the numerical constant $C$, we get

\begin{eqnarray}
	\omega_D = -(H_0^2 + k)\left[\frac{\alpha}{k + H_0^2(1 - \Omega_{m0})} + \frac{\beta}{H_0^2\Omega_{m0}}\right]\label{wD}.
\end{eqnarray}

\hspace{0.5cm} From (\ref{diffFriedmann}),(\ref{C}) and (\ref{wD}) can help us to find other cosmological parameters, first for the Hubble parameter can be written as

\begin{eqnarray}
	H = \sqrt{-\frac{k}{a^2} + a^{\frac{6\left[H_0^2(1 - \alpha + \beta)\Omega_{m0} - (H_0^2 + k)\right]}{(\Delta - 2)H_0^2\Omega_{m0}}}{H_0^2 + k}}\label{FinalHubble}.
\end{eqnarray}

\hspace{0.5cm} Furthermore, for the deceleration parameter, where $q = -1 -\frac{aH^\prime}{H}$, we are able to find this parameter as an analytic solution by using (\ref{diffFriedmann}) as

\begin{eqnarray}
	q = \frac{(1 + \Delta + 3\omega_D)\Omega_D + (1 + \Delta)\Omega_m}{2 - \Delta}\label{q:analytic}
\end{eqnarray}
	\section{Results and Discussion}
\hspace{0.5cm} We use the initial data from the Planck base-$\Lambda$CDM results \cite{aghanim2020planck} or particularly we use $H_0=64.7$ km/s/Mpc, $\Omega_{m0}=0.315$ and $\Omega_{k0}=0.0007$ for the Hubble parameter, the matter density parameter and the curvature density parameter in present cosmic time, respectively. We use the coupling parameters $\alpha^2$ and $\beta^2$ with values of 0.0225 and 0.0625, respectively.
\begin{figure}[h]
	\centering
	\subfloat[Both positive value of $\alpha$ and $\beta$]{\includegraphics[width = 0.5\textwidth]{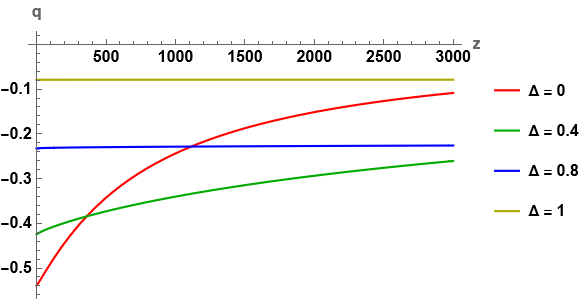}\label{fig:qapbp}}
	\hfill
	\subfloat[Both negative value of $\alpha$ and $\beta$]{\includegraphics[width = 0.5\textwidth]{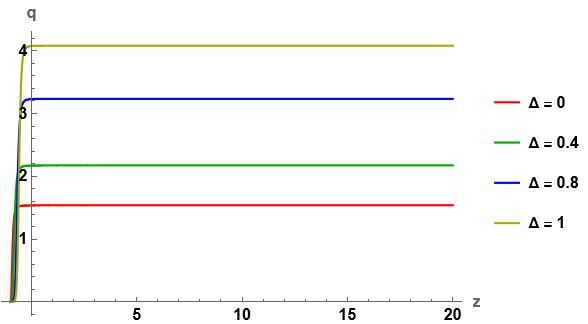}\label{fig:qambm}}
	
	\subfloat[Positive $\alpha$ and negative $\beta$]{\includegraphics[width = 0.5\textwidth]{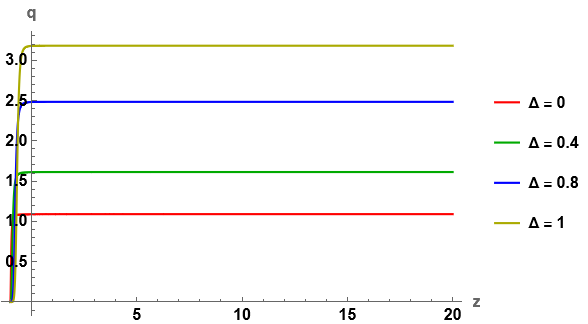}\label{fig:qapbm}}
	\hfill
	\subfloat[Negative $\alpha$ and positve $\beta$]{\includegraphics[width = 0.5\textwidth]{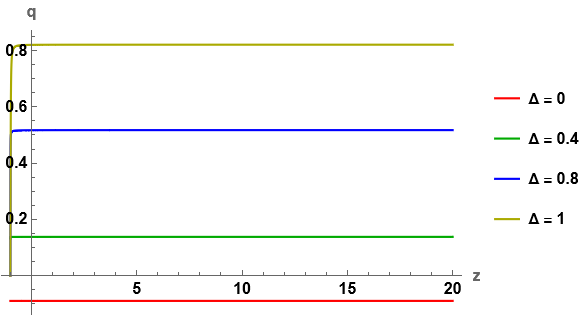}\label{fig:qambp}}
	\caption{Plots of the deceleration parameter $q$ versus the redshift parameter $z$ in respective values of coupling parameter $\alpha$ and $\beta$.}
	\label{fig:deceleration}
\end{figure}

\hspace{0.5cm} From \eqref{wD} and \eqref{q:analytic}, it is obvious that when there is no dark sector interaction $(\alpha=\beta=0)$, the equation of state parameter $\omega_D$ is always zero and the deceleration parameter $q$ is always positive. Thus, the thermodynamically satisfied BHDE model using the apparent horizon as the IR cut-off without any dark sector interaction does not result in the accelerating expansion of the universe. For the linearly interacting model case, from Figure \ref{fig:deceleration}, we have negative $q$ for every case of different values of $\Delta$ when the coupling parameters $\alpha$ and $\beta$ are both positive. This means, in the model with the linear interaction, the acceleration problem has been solved despite the fact that we did not find any phase transition of the universe expansion from deceleration $(q>0)$ to acceleration $(q<0)$.

\begin{figure}[h]
	\centering
	\subfloat[Both positive value of $\alpha$ and $\beta$]{\includegraphics[width = 0.5\textwidth]{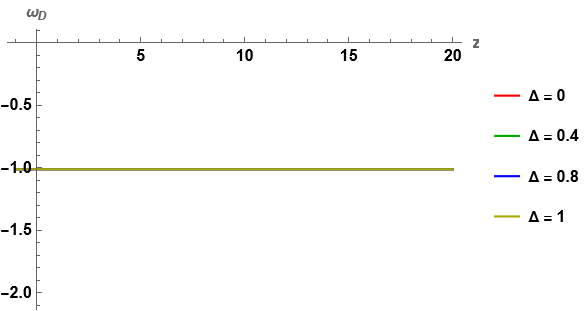}\label{fig:wdapbp}}
	\hfill
	\subfloat[Both negative value of $\alpha$ and $\beta$]{\includegraphics[width = 0.5\textwidth]{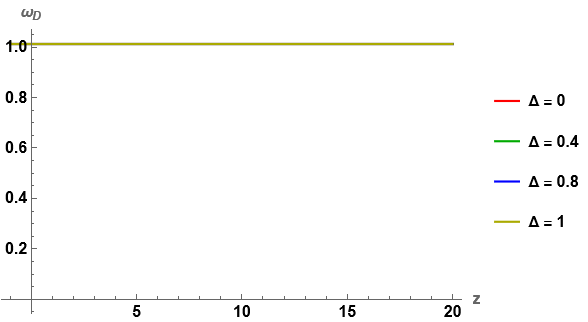}\label{fig:wdambm}}
	
	\subfloat[Positive $\alpha$ and negative $\beta$]{\includegraphics[width = 0.5\textwidth]{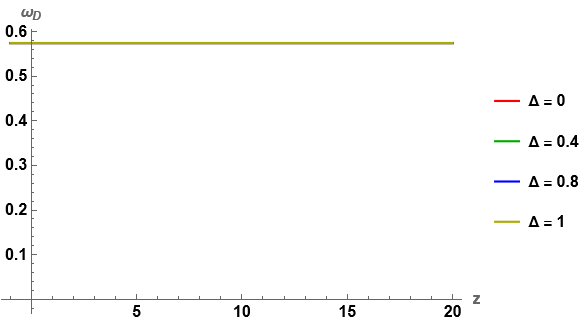}\label{fig:wdapbpm}}
	\hfill
	\subfloat[Negative $\alpha$ and positive $\beta$]{\includegraphics[width = 0.5\textwidth]{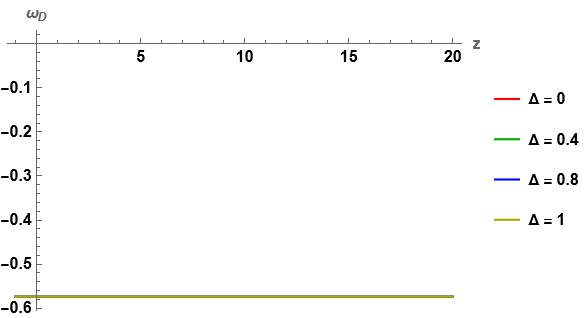}}
	\caption{Plots of the equation of state parameter $\omega_D$ versus the redshift parameter $z$ in respective values of coupling parameter $\alpha$ and $\beta$.}
	\label{fig:w}
\end{figure}

\hspace{0.5cm} For the equation of state parameter in Figure \ref{fig:w}, we have the dark energy phantom phase for the case of the positive values of $\alpha$ and $\beta$. This satisfies the observational data from \cite{aghanim2020planck}. For the case of the negative values of $\alpha$ and $\beta$, we have the dark energy in the quintessence phase.

\begin{figure}[h]
	\centering
	\subfloat[Both positive value of $\alpha$ and $\beta$]{\includegraphics[width = 0.5\textwidth]{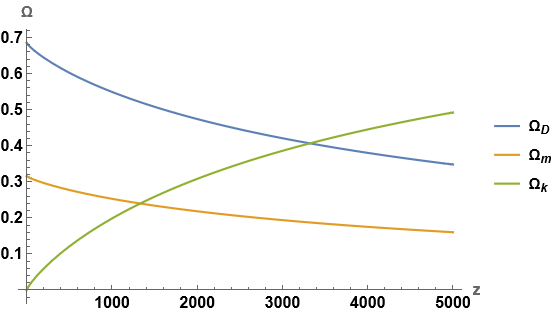}\label{fig:Omegaapbp}}
	\hfill
	\subfloat[Both negative value of $\alpha$ and $\beta$]{\includegraphics[width = 0.5\textwidth]{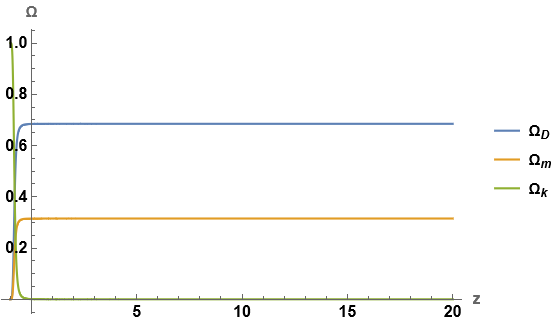}\label{fig:Omegaambm}}
	
	\subfloat[Positive $\alpha$ and negative $\beta$]{\includegraphics[width = 0.5\textwidth]{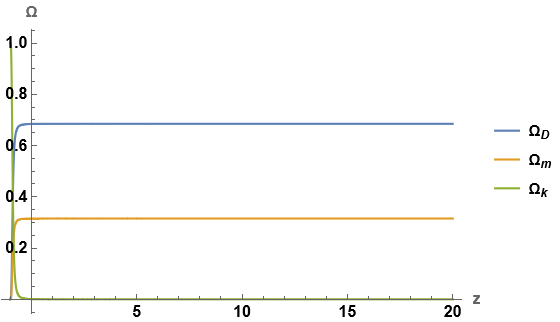}\label{fig:Omegaapbm}}
	\hfill
	\subfloat[Negative $\alpha$ and positive $\beta$]{\includegraphics[width = 0.5\textwidth]{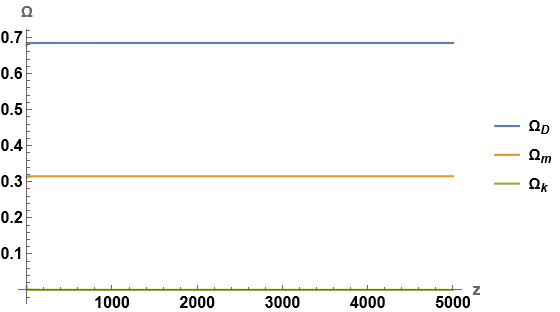}\label{fig:Omegaambp}}
	
	\caption{Plots of the density parameter $\Omega$ versus the redshift parameter $z$ in respective values of coupling parameter $\alpha$ and $\beta$.}
	\label{fig:W}
\end{figure}

\hspace{0.5cm} The results illustrated in Figure \ref{fig:W} do not agree with the observable data, which indicates that matter dominated in the previous era. Figure \ref{fig:W} shows that dark energy has dominated throughout all of cosmic evolution except for the case where the coupling parameters are both positive, as shown in Figure \ref{fig:Omegaapbp}. For this figure, there is an interesting behaviour of the density parameter for the curvature component $\Omega_k$, which is not fine-tuned yet dominates in the early universe and decays to a flat geometry universe at present, alleviating the flatness problem. Unfortunately, this model does not agree with the result from the $\Lambda$CDM model; the matter component of the linear interaction model has not dominated during cosmic evolution. Nevertheless, by using \eqref{BHDE}, \eqref{apparent horizon}, \eqref{OmegaAll}, \eqref{the density parameters} and \eqref{FinalHubble}, we have
\begin{equation}
	\frac{\Omega_D}{\Omega_m}=\frac{\Omega_{D0}}{\Omega_{m0}}
\end{equation}

so that this model leaves the question of the coincidence problem only to the early phase of the universe, i.e. in this model the ratio between the dark energy density parameter and the matter density parameter remains constant at all time, and this constant is set by the initial condition of the universe.

\begin{figure}[h]
	\centering
	\subfloat[Both positive value of $\alpha$ and $\beta$]{\includegraphics[width = 0.5\textwidth]{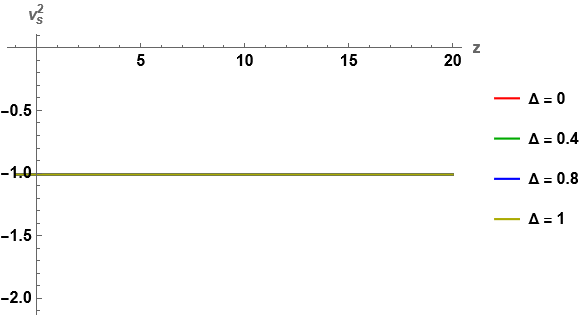}\label{fig:vsqapbp}}
	\hfill
	\subfloat[Both negative value of $\alpha$ and $\beta$]{\includegraphics[width = 0.5\textwidth]{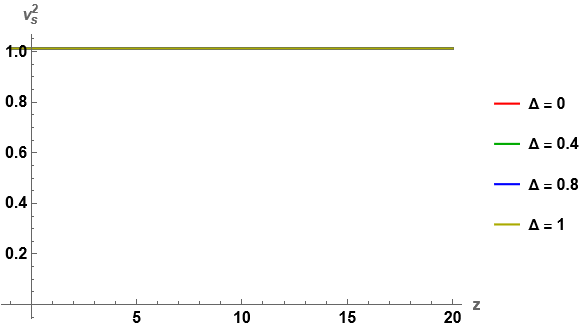}\label{fig:vsqambm}}
	
	\subfloat[Positive $\alpha$ and negative $\beta$]{\includegraphics[width = 0.5\textwidth]{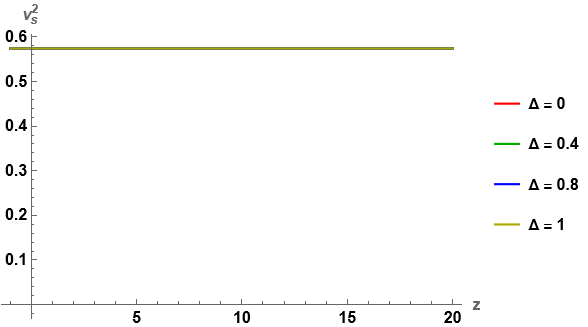}\label{fig:vsqapbm}}
	\hfill
	\subfloat[Negative $\alpha$ and positive $\beta$]{\includegraphics[width = 0.5\textwidth]{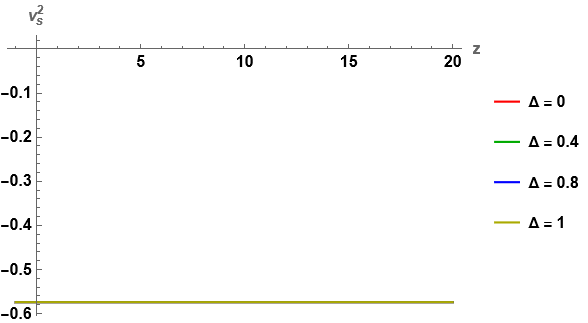}\label{fig:vsqambp}}	
	\caption{Plots of the sound speed squared parameter $v_s^2$ versus the redshift parameter $z$ in respective values of coupling parameter $\alpha$ and $\beta$.}
	\label{fig:vs}
\end{figure}

\hspace{0.5cm} For the instability analysis, we consider the sound speed squared of dark energy \cite{peebles2003cosmological}

\begin{eqnarray}
	v_s^2 = \frac{dp_D}{d\rho_D},
\end{eqnarray}

by using the equation of state parameter definition, we obtain

\begin{eqnarray}
	v_s^2 = \frac{d(\omega_D\rho_D)}{da} = \omega_D + \frac{\omega_D^\prime}{\rho_D^\prime}.
\end{eqnarray}

\hspace{0.5cm} Together with \eqref{wD}, we obtain the sound speed squared having the same expression as the equation of state parameter, i.e. $v_s^2=\omega_D$. The stable range of sound speed squared lies in $0 \leq v_s^2 \leq 1$. However, from Figure \ref{fig:vs}, for $0 \leq \Delta \leq 1$, in the case of the universe with accelerated expansion, we have negative values of the sound speed squared. This means the linearly interacting BHDE model with apparent horizon as the IR cut-off is unstable during the cosmic evolution where the universe has accelerated expansion.
\section{Conclusion}
\hspace{0.5cm} The non-interacting thermodynamically satisfied Barrow HDE model where the IR cut-off is the apparent horizon has the acceleration problem. The model with the general linear interaction between dark sectors solves this, and has the dark energy in phantom phase satisfied with the observation\cite{aghanim2020planck}. This model also alleviates the flatness and coincidence problems. However, there are some difficulties arised in the model. First, we found that the deceleration parameter presents no phase transition from the decelerating to accelerating expansion of the universe. There is also instability from the negativity of the sound speed squared parameter, though we varied different values of the Barrow exponent $\Delta$. Moreover, the density parameter of the matter component does not correspond to $\Lambda$CDM, as this parameter has to dominate in the past. However, in the model we study, we found, in the situation that universe has accelerated expansion, that the curvature density parameter instead dominated in the past. Nonetheless, as the universe evolve to the future, the curvature density parameter decays to the flat-spatial universe. The next problem is that we found the dark energy equation of state parameter is the same as the sound speed squared, so this means that this model is not able to achieve stability while the dark energy is lying in the quintessence or phantom regime. We hope our future work, which involves changing the constant coupling parameters to the dynamical coupling parameter will alleviate the phase transition, instability and matter domination problems.

	\newpage
\section{Acknowledgments}
\hspace{0.5cm} We thank the 19th Siam Physics Congress 2024 (SPC2024) and its organisers and commitees for some good suggestions. KD also thanks Amonthep Tita for some useful discussion about HDE models in other perspectives.
	\section{References}
\renewcommand{\refname}{}
\bibliographystyle{unsrt}
\bibliography{References}

\end{document}